\documentclass[aps,twocolumn,
superscriptaddress,
footinbib,
noeprint,
pra]{revtex4-2}

\usepackage[utf8]{inputenc} 
\usepackage{mathtools,amssymb,bm}
\usepackage{graphicx}
\usepackage{epstopdf}
\usepackage{latexsym}
\usepackage[normalem]{ulem} 
\usepackage[caption=false, position=top]{subfig}
\usepackage[usenames,dvipsnames]{color}
\usepackage{hyperref}
\usepackage{xcolor}
\usepackage{physics}
\usepackage{multirow}
\usepackage{stackengine}
\usepackage{dsfont}
\usepackage{bbold}
\usepackage{nicefrac}
\usepackage[inkscapelatex=false]{svg}
\usepackage{enumitem}

\usepackage{framed}
\graphicspath{{gfx/}}

\newcommand{\ind}[1]{_\text{#1}}

\newcommand{\comment}[1]{}

\def\equationautorefname~#1\null{Equation~(#1)\null}
\begin{document}

\title{Pair localization in dipolar systems with tunable positional disorder}

\date{\today}

\author{Adrian Braemer}
\email{adrian.braemer@kip.uni-heidelberg.de}
\affiliation{Physikalisches Institut, Universität Heidelberg, Im Neuenheimer Feld 226, 69120 Heidelberg, Germany}
\author{Titus Franz}
\affiliation{Physikalisches Institut, Universität Heidelberg, Im Neuenheimer Feld 226, 69120 Heidelberg, Germany}
\author{Matthias Weidemüller}
\affiliation{Physikalisches Institut, Universität Heidelberg, Im Neuenheimer Feld 226, 69120 Heidelberg, Germany}
\author{Martin Gärttner}
\email{martin.gaerttner@kip.uni-heidelberg.de}
\affiliation{Physikalisches Institut, Universität Heidelberg, Im Neuenheimer Feld 226, 69120 Heidelberg, Germany}
\affiliation{Kirchhoff-Institut für Physik, Universität Heidelberg, Im Neuenheimer Feld 227, 69120 Heidelberg, Germany}
\affiliation{Institut für Theoretische Physik, Ruprecht-Karls-Universität Heidelberg, Philosophenweg 16, 69120 Heidelberg, Germany}

\begin{abstract}
Strongly interacting quantum systems subject to quenched disorder exhibit intriguing phenomena such as glassiness and many-body localization. Theoretical studies have mainly focused on disorder in the form of random potentials, while many experimental realizations naturally feature disorder in the interparticle interactions. 
Inspired by cold Rydberg gases, where such disorder can be engineered using the dipole blockade effect,
we study a Heisenberg XXZ spin model where the disorder is exclusively due to random spin-spin couplings, arising from power-law interactions between randomly positioned spins.
Using established spectral and eigenstate properties and entanglement entropy, we show that this system exhibits a localization crossover and identify strongly interacting pairs as emergent local conserved quantities in the system, leading to an intuitive physical picture consistent with our numerical results.
\end{abstract}

\maketitle  

\section{Introduction}

Understanding how an isolated quantum system prepared out of equilibrium, can exhibit thermal properties at late times, i.e., how it thermalizes, has challenged quantum physicists for almost a century. 
The eigenstate thermalization hypothesis (ETH) \cite{srednickiChaosQuantumThermalization1994,deutschEigenstateThermalizationHypothesis2018} offers a generic mechanism to explain this phenomenon but makes strong assumptions on the structure of energy eigenstates in terms of the matrix elements of local operators. 
Nonetheless, it has been shown numerically that a large class of quantum systems complies with ETH and thermalizes \cite{gogolinEquilibrationThermalisationEmergence2016,dalessioQuantumChaosEigenstate2016}. 
A notable exception are strongly disordered systems in which transport is absent and the system retains memory of the initial state at arbitrary times \cite{nandkishoreManyBodyLocalizationThermalization2015,abaninRecentProgressManybody2017a,nandkishoreManyBodyLocalization2017,abaninManybodyLocalizationThermalization2019}. 

This phenomenon, called many-body localization (MBL), has been verified for small systems including, but not limited to, spin systems with random potentials \cite{znidaricManyBodyLocalization2008,luitzManybodyLocalizationEdge2015,sierantPolynomiallyFilteredExact2020}, random nearest \cite{vasseurParticleholeSymmetryManybody2016,protopopovNonAbelianSymmetriesDisorder2020,chandaManybodyLocalizationTransition2020} and next-to-nearest neighbor interactions \cite{kjallManyBodyLocalizationDisordered2014,bahovadinovManybodyLocalizationTransition2022}, and power-law interactions \cite{burinManybodyDelocalizationStrongly2015,schifferManybodyLocalizationSpin2019,roySelfconsistentTheoryManybody2019,safavi-nainiQuantumDynamicsDisordered2019,mohdebExcitedEigenstateEntanglementProperties2022} using a combination of exact numerical approaches and heuristic arguments like the strong disorder renormalization group (SDRG) \cite{pekkerHilbertGlassTransitionNew2014,potterUniversalPropertiesManybody2015,voskTheoryManyBodyLocalization2015,monthusStrongDisorderRenormalization2018} to generalize to large systems. 

Recently, claims have been made that this localization phenomenology may not be stable in the thermodynamic limit due to thermal inclusions \cite{deroeckStabilityInstabilityDelocalization2017,luitzHowSmallQuantum2017,morningstarAvalanchesManybodyResonances2022,selsDynamicalObstructionLocalization2021,selsThermalizationDiluteImpurities2022,selsBathinducedDelocalizationInteracting2022,thieryManyBodyDelocalizationQuantum2018,deroeckStabilityInstabilityDelocalization2017,ponteThermalInclusionsHow2017,pandeyAdiabaticEigenstateDeformations2020}. These are small, more ordered subregions thought to thermalize with their surroundings and thus slowly pushing the system toward thermalization. Unfortunately, these regions are very rare and thus only start appearing in large systems far beyond the reach of numerical methods.
This raises the question, whether this instability is relevant for quantum simulation experiments, being finite in size and limited by coherence time. In this paper, we only focus on the phenomenology of localization in finite systems and subsequently use the term localized regime instead of a phase, following the terminology of Ref. \cite{morningstarAvalanchesManybodyResonances2022}.

Complementary to numerical works, there are a number of experimental results falling into roughly two classes: Experiments with single particle resolution, including optical lattices \cite{kondovDisorderInducedLocalizationStrongly2015,schreiberObservationManybodyLocalization2015,bordiaCouplingIdenticalOnedimensional2016,lukinProbingEntanglementManybody2019} and trapped ions \cite{smithManybodyLocalizationQuantum2016}, and experiments based on macroscopic samples, like NV centers in diamond \cite{kucskoCriticalThermalizationDisordered2018} or NMR systems \cite{weiExploringLocalizationNuclear2018}. The former offer precise control, but are rather limited in size, while the latter can realize much larger systems at the expense of flexibility, in particular, lack of programmable disorder.
Cold gases of Rydberg atoms implement dipolar dynamics with random couplings (similar to NMR systems or NV centers) and allow for control of the disorder strength and even the power law of the interaction at rather large particle numbers \cite{signolesGlassyDynamicsDisordered2021}, which makes them a powerful platform for studying localization phenomena.

Motivated by recent progress on quantum simulations with Rydberg atoms \cite{orioliRelaxationIsolatedDipolarInteracting2018,signolesGlassyDynamicsDisordered2021,geierFloquetHamiltonianEngineering2021,franzAbsenceThermalizationInteracting2022}, we consider a power-law interacting spin system where the disorder is due to randomly positioned spins respecting a blockade condition, which induces disordered couplings.
In this setup, the strength of the disorder can be tuned by changing the density of particles or, equivalently, the minimal distance between them. Starting in a ordered system, where the blockade radius is of order of the mean inter-particle distance, we show numerically that this system exhibits a crossover to a localized regime at small blockade and apply a SDRG approach to derive a simple model based on strongly interacting pairs, which captures the properties of the eigenstates in the localized regime well. Our study thus adds to the body of numerical works on MBL, focusing on dipolar systems with tunable positional disorder, and is highly relevant to experimental efforts, as a wide range of quantum simulation platforms feature dipolar interactions.

\section{Localization in a Rydberg gas}

\subsection{System}
We consider the Heisenberg XXZ spin model described by the Hamiltonian ($\hbar = 1$)
\begin{equation}
	\hat{H} = \frac{1}{2}\sum_{i\neq j} J_{ij} \underbrace{ \left( \hat{S}_x^{(i)}\hat{S}_x^{(j)} + \hat{S}_y^{(i)} \hat{S}_y^{(j)} + \Delta \hat{S}_z^{(i)} \hat{S}_z^{(j)} \right) }_{\equiv H\ind{pair}^{(i)(j)}}  \,
	\label{eq:H_XXZ}
\end{equation}
where $\hat{S}_{\alpha}^{(k)}$ (with $\alpha \in \{x,y,z\}$) denotes the spin-$\frac{1}{2}$ operators acting on the $k$-th spin. The coupling $J_{ij}$ between spins $i$ and $j$ at positions $x_i$ and $x_j$ is given by $J_{ij}=\frac{C_\alpha}{|x_i-x_j|^\alpha}$, where $C_\alpha$ is an interaction coefficient which we set to $C_\alpha=1$. 
In experimental realizations of this model with Rydberg atoms, the values of the anisotropy parameter $\Delta$ and interaction exponent $\alpha$ are controllable via the choice of the Rydberg states encoding the two spin states. The cases $\alpha=3$, $\Delta=0$ (dipolar exchange) and $\alpha=6$, $\Delta\approx -0.7$ (van-der-Waals) have been realized experimentally \cite{signolesGlassyDynamicsDisordered2021,geierFloquetHamiltonianEngineering2021}. For typical cloud temperatures and time scales of the spin dynamics the atom positions can be regarded as fixed (frozen gas approximation).

During the initial Rydberg excitation, the spins are subjected to the Rydberg blockade \cite{lukinDipoleBlockadeQuantum2001} which means no two spins can be closer than some distance $r_b$, called the blockade radius. This feature allows one to tune the strength of disorder via the sample's density: In a very dilute sample, the mean inter-spin distance is much larger than the blockade radius $r_b$ and thus positions are essentially uncorrelated. In the other extreme, the spins are tightly packed and exhibit strong spatial correlation.

We quantify the strength of disorder by the ratio $W$ of the system's total volume $V$ over total blocked volume $V\ind{block}$ or, equivalently, by the ratio of Wigner-Seitz radius $a_0$, which is half of the mean inter-spin distance, to the blockade radius $r_b$ to the power of the dimension $d$:
\begin{equation}
    W = \frac{V}{V\ind{block}} = \left( \frac{a_0}{r_b} \right)^d \, .
\end{equation}
For $d=1$, the minimal value of $W_{min}=\frac{1}{2}$ is attained for a translationally invariant chain with spacing $2a_0 = r_b$, as illustrated in Fig.~\ref{fig:1}(a).

\subsection{Effective pair description}
\label{sec:pair_model}
\begin{figure}
    \centering
    \def\svgwidth{0.45\textwidth}
\begingroup%
  \makeatletter%
  \providecommand\color[2][]{%
    \errmessage{(Inkscape) Color is used for the text in Inkscape, but the package 'color.sty' is not loaded}%
    \renewcommand\color[2][]{}%
  }%
  \providecommand\transparent[1]{%
    \errmessage{(Inkscape) Transparency is used (non-zero) for the text in Inkscape, but the package 'transparent.sty' is not loaded}%
    \renewcommand\transparent[1]{}%
  }%
  \providecommand\rotatebox[2]{#2}%
  \newcommand*\fsize{\dimexpr\f@size pt\relax}%
  \newcommand*\lineheight[1]{\fontsize{\fsize}{#1\fsize}\selectfont}%
  \ifx\svgwidth\undefined%
    \setlength{\unitlength}{312.36700814bp}%
    \ifx\svgscale\undefined%
      \relax%
    \else%
      \setlength{\unitlength}{\unitlength * \real{\svgscale}}%
    \fi%
  \else%
    \setlength{\unitlength}{\svgwidth}%
  \fi%
  \global\let\svgwidth\undefined%
  \global\let\svgscale\undefined%
  \makeatother%
  \begin{picture}(1,0.88741254)%
    \lineheight{1}%
    \setlength\tabcolsep{0pt}%
    \put(0,0){\includegraphics[width=\unitlength,page=1]{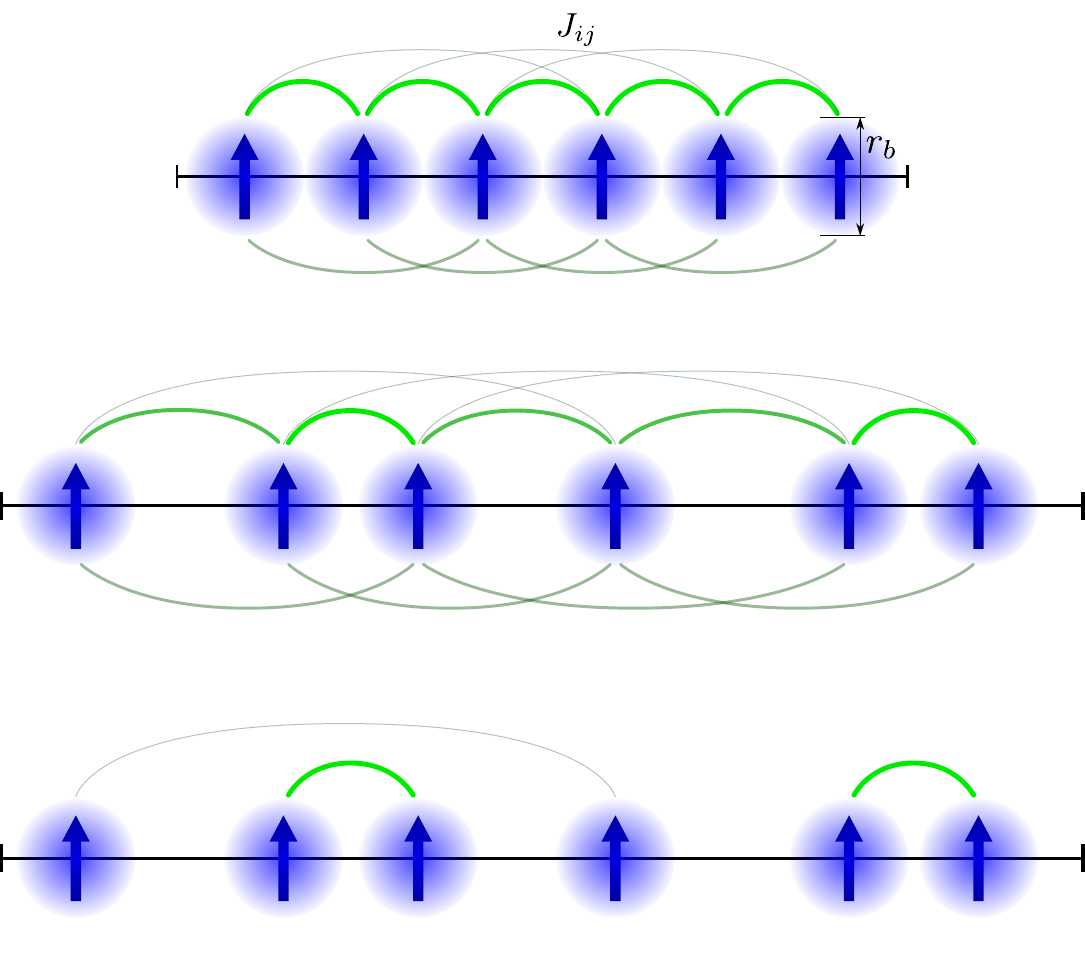}}%
    \put(-0.00220094,0.86795421){\makebox(0,0)[lt]{\lineheight{1.25}\smash{\begin{tabular}[t]{l}(a) fully ordered\end{tabular}}}}%
    \put(-0.00220094,0.59063552){\makebox(0,0)[lt]{\lineheight{1.25}\smash{\begin{tabular}[t]{l}(b) disordered\end{tabular}}}}%
    \put(-0.00220094,0.25584194){\makebox(0,0)[lt]{\lineheight{1.25}\smash{\begin{tabular}[t]{l}(c) pair description\end{tabular}}}}%
    \put(0,0){\includegraphics[width=\unitlength,page=2]{figure1.pdf}}%
    \put(0.29141396,0.01745787){\makebox(0,0)[lt]{\lineheight{1.25}\smash{\begin{tabular}[t]{l}LIOM\end{tabular}}}}%
  \end{picture}%
\endgroup%

    \caption{\textbf{Pair description.} The blockade constraint (blue shadings) enables tuning of disorder in the couplings (green lines) from fully ordered (a) to disordered (b). In the latter case a perturbative treatment to first order yields a description in terms of strongly correlated pairs (c) subject to an Ising-like interaction (not depicted). These pairs constitute local integrals of motion (LIOM).}
    \label{fig:1}
\end{figure}

This model differs from the random field Heisenberg model, which has been studied extensively in the MBL literature, as no disordered potentials are considered. Thus it may not be immediately apparent, why this system features localization and what constitutes the local conserved quantities akin to the $l$-bits \cite{husePhenomenologyFullyManybodylocalized2014} in the standard scenario. Here we provide a phenomenological picture in the spirit of the SDRG suggesting that localization should appear due to strongly interacting pairs.

Consider a strongly disordered cloud of $N$ spins described by Eq.~\eqref{eq:H_XXZ} like the example depicted in Fig.~\ref{fig:1}(b). Due to the power-law interactions, coupling strengths vary strongly between different pairs of atoms, symbolized by the width and brightness of the green lines. This motivates us to employ a perturbative treatment, in which we single out the strongest pair coupling and consider all other couplings as a perturbation. In the example shown in  Fig.~\ref{fig:1}(b), the two rightmost spins share the strongest coupling and we can see that it is much stronger than the other couplings of either one of the spins to the rest of the system. Using perturbation theory to first order, we find  that the pair of spins almost decouples from the rest of the system leaving only an effective Ising-like interaction, which is unimportant for the further procedure and thus not shown in the figure. For details on the calculations involved, see Appendix \ref{appendix:pair_model}.

We may now repeat this procedure of eliminating couplings between the pairs and the rest of system by identifying the next strongest interaction among the remaining spins which, in this example, is the coupling between the second and third spin. Eliminating the respective couplings as well leaves us with the effective pairs shown in Fig.~\ref{fig:1}(c). Note that in an ordered system, as shown in Fig.~\ref{fig:1}(a), this perturbative treatment is not applicable as not all neglected couplings can be considered small. We also note that the order of eliminations is not important as long as each time the inner-pair coupling is much larger than the couplings between the pair and the rest. Concretely, for the given example, choosing the coupling between spins 2 and 3 in Fig.~\ref{fig:1}(b) first in the pair elimination process does not change the result. 

The great advantage of this ansatz is that we can now give a simple description of the whole many-body spectrum. Diagonalizing $H\ind{pair}$ [see Eq ~\ref{eq:H_XXZ}], we find two maximally entangled eigenstates $\ket{\pm} = 1/\sqrt{2} (\ket{\uparrow\downarrow} \pm \ket{\downarrow\uparrow})$ at energies $E_\pm = \pm 2 - \Delta$ and two degenerate states $\ket{\uparrow\uparrow}$, $\ket{\downarrow\downarrow}$ at energy $E_d = \Delta$, which we will refer to as $\ket{\updownarrow\updownarrow}$. The Ising-like interaction between pairs does not act on the entangled states $\ket{\pm}$ and is diagonal with respect to $\ket{\updownarrow\updownarrow}$. Thus, in the pair picture, the eigenstates of the full system are now given by tensor products of these four pair eigenstates. We refer to this basis as the pair basis.

In the many-body spectrum, the degeneracy between the pair states $\ket{\uparrow\uparrow}$ and $\ket{\downarrow\downarrow}$ is lifted due to the emerging Ising-like interaction.
However, we note that this splitting is small compared to the splitting between the other pair eigenstates as it emerges from first order perturbation theory.

The pair picture is analogous to the $l$-bit picture often used in MBL, where strong local disorder potentials lead to the emergence of quasi-local conserved quantities $\hat{\tau}^{(i)}\sim \hat{\sigma}_z^{(i)}$ \cite{husePhenomenologyFullyManybodylocalized2014,serbynLocalConservationLaws2013}. Here, we see that each projector on a pair's eigenstate constitutes an approximately conserved quantity and hence is a local integral of motion (LIOM). Thus, we established a description akin to the $l$-bit picture of MBL for this disordered Heisenberg model, where the role of LIOMs is taken by strongly interacting pairs.

While this ansatz is heuristic and neglects all higher resonances, that may play a crucial role in delocalizing the system, it will nonetheless turn out to be useful for interpreting and understanding the spectral and eigenstate properties reported in the following.

\section{Numerical Results}
To minimize boundary effects, we consider a one\nobreakdash-dimensional system with periodic boundary conditions \footnote{Only the closest copy of each spin is considered for the interaction.} of up to $N=16$ spins governed by Eq.~\eqref{eq:H_XXZ} and perform exact diagonalisation on the sector of smallest positive magnetization. We fix the interaction exponent to $\alpha = 6$, corresponding to a Van-der-Waals interactions, and set $\Delta=-0.73$ (cf. Ref. \cite{signolesGlassyDynamicsDisordered2021}). We do not expect a strong dependence of our results on the precise value of $\Delta$ as long as one steers clear from regions around points where additional symmetries emerge.

For each disorder strength $W$, we generate 2000 configurations, perform a full diagonalization and compute several well-established indicators for the localization transition from the spectrum. We always average over all eigenstates/-values as restricting to the bulk of the spectrum does not lead to qualitative changes in the observed behavior. The statistical error resulting from disorder averaging is smaller than the thickness of the lines unless indicated otherwise. For a description of the algorithm for choosing the configurations, we refer to Appendix~\ref{appendix:drawing_positions}. All code used for this paper can be found in Ref.~\footnote{\url{https://github.com/abraemer/Pair-localization-paper}}.

The following sections discuss different indicators of localization with the aim to establish the localization crossover in this model and employ the pair model for interpretation and predictions. The last section directly compares the pair basis to the eigenstates, thus demonstrating its validity.

\subsection{Level spacing ratio}
The spectral average of the level spacing ratio (LSR), defined as \cite{oganesyanLocalizationInteractingFermions2007} 
\begin{equation}
\langle r \rangle = \frac{1}{|\mathcal{H}|}\sum_n \min\left(\frac{E_{n+2} - E_{n+1}}{E_{n+1} - E_{n}}, \frac{E_{n+1} - E_{n}}{E_{n+2} - E_{n+1}}\right),
\end{equation}
is a simple way of characterizing the distribution of differences between adjacent energy levels. For thermalizing (ergodic) systems, the Hamiltonian is expected to show a mean LSR resembling a random matrix from the Gaussian orthogonal ensemble because its eigenvectors essentially look like random vectors. Thus one can use random matrix theory to obtain $\langle r \rangle\ind{thermal} = 4 - 2\sqrt{3} \approx 0.536 $ \cite{atasDistributionRatioConsecutive2013}.

On the other hand, in localized systems the eigenvalues follow a Poisson distribution, since they are essentially sums of randomly distributed energies from the $l$-bits the system consists of. Computing the mean LSR in this case yields $\langle r \rangle\ind{MBL} = 2 \ln 2 - 1 \approx 0.386 $ \cite{atasDistributionRatioConsecutive2013}.

\begin{figure}
    \includegraphics[width=0.45\textwidth]{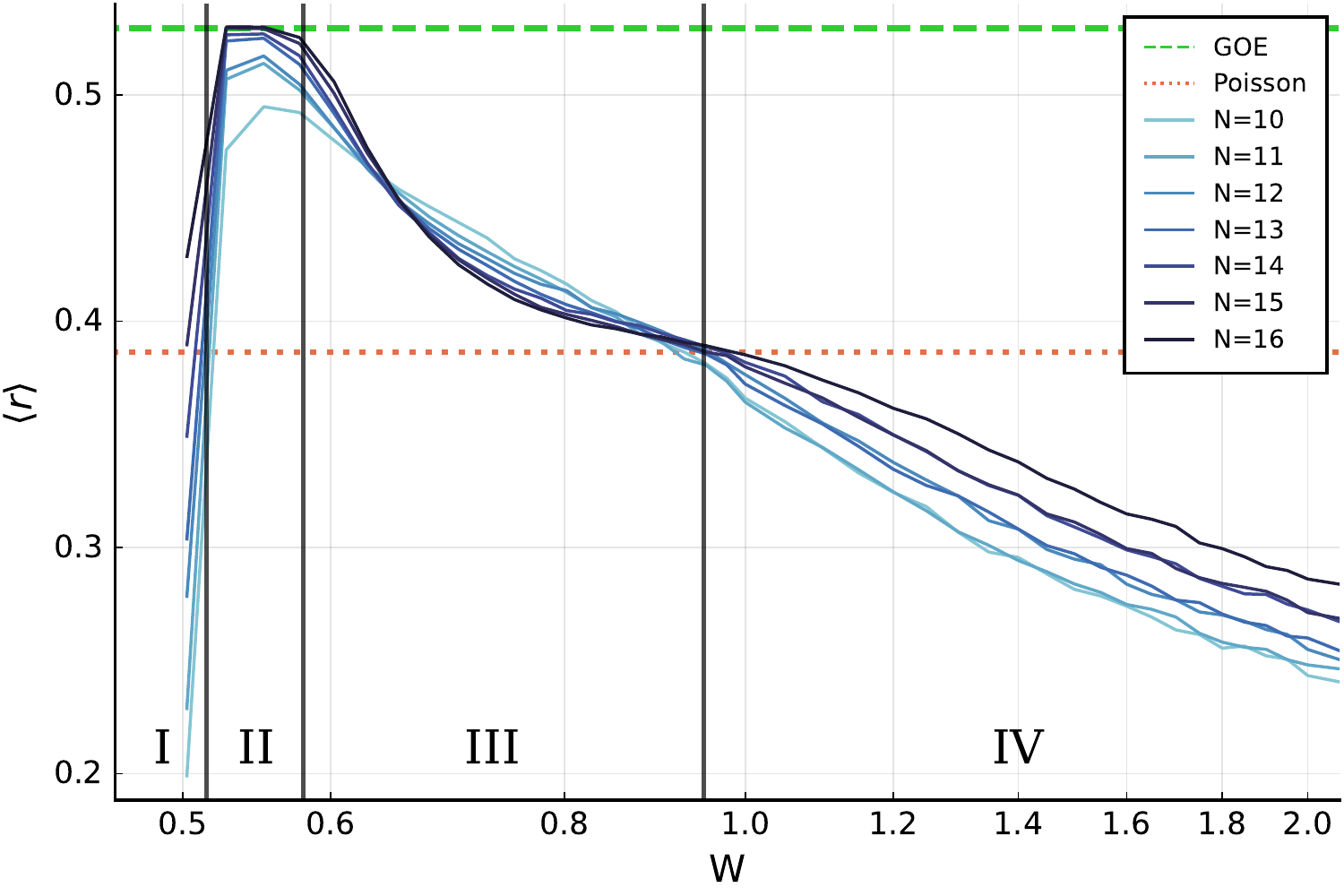}
    \caption{\textbf{Level-spacing ratio.} With increasing disorder the LSR shows a crossover from an ergodic value to its Poissonian value and below. We identify four major regions where the physics is governed by (I) translational symmetry breaking, (II) thermal behavior, (III) the localization crossover and (IV) localization. The horizontal lines show random-matrix theory predictions.}
    \label{fig:lsr}
\end{figure}

Comparing with the numerical results in Fig.~\ref{fig:lsr} and focusing on the central parts first, we find the mean LSR reaches its thermal value for large enough systems and weak disorder (II) dropping towards the Poissonian value for stronger disorder (III). With growing system size, the thermal plateau (II) broadens, marking a parameter region where the system appears ergodic. But while the plateau broadens, the drop-off (III) for increasing disorder strength becomes steeper, meaning the crossover becomes sharper as the system gets larger.

Considering very strong disorder (IV), the mean LSR drops even below the Poissonian value, which indicates level attraction. This effect can be explained by the pair model: As stated earlier, the $\ket{\updownarrow\updownarrow}$ states' degeneracy is lifted by the effective Ising-like terms from first-order perturbation theory, which means the split is of smaller magnitude compared to the intra-pair interactions. For small systems with comparatively low spectral density, this means that the small lifting likely fails to mix the formerly degenerate states into their surrounding spectrum. Thus, the LSR still reflects the near degeneracy within the pairs, leading to level attraction. Based on this interpretation, we expect this effect to diminish for larger systems with the spectral density growing. In fact, this trend is already visible in Fig.~\ref{fig:lsr}.

A similar argument can be made at very weak disorder (I): Here the source of the degeneracy is the proximity to the perfectly ordered case at $W=0.5$, which has an additional translation invariance. Weak disorder breaks that symmetry but couples the symmetry sectors only weakly, leading again to a very small energetic splitting of degenerate states. We want to emphasize the reason for level attraction being very different in nature in (I) and (IV): Whereas in (I) the system is close to a system with obvious conserved quantities due to symmetries, in (IV) there is the emergent integrability of the MBL regime\cite{abaninManybodyLocalizationThermalization2019}. Nevertheless, we expect region (I) to become less pronounced for larger systems continuing the trend visible in Fig.~\ref{fig:lsr}.

We conclude, that, in analogy to standard MBL, we find a crossover in the level spacing distribution from a regime with level repulsion to Poissonian gaps indicating a localization crossover. At very strong disorder, we even find a region with level attraction, the source of which can be explained by the effective pair model.

\subsection{Thouless parameter}
Complementary to eigenvalue statistics, we also probe eigenstate properties by computing the Thouless parameter
\begin{equation}
    \mathcal{G}_n = \ln \frac{|\mel{n}{\hat{V}}{n+1}|}{E^\prime_{n+1} - E^\prime_n},
\end{equation}
introduced by \citeauthor{serbynCriterionManybodyLocalizationdelocalization2015}\cite{serbynCriterionManybodyLocalizationdelocalization2015}. This quantity is akin to the Thouless conductance in single particle systems and quantifies how well two states $\ket{n}$,$\ket{n+1}$ with perturbed energies $E^\prime_n = E_n + \expval{\hat{V}}{n}$ are coupled by a local perturbation $\hat{V}$. In the thermal phase, states of similar energy will have similar spatial structures, whereas in the localized phase, eigenstates are products of LIOM eigenstates and thus typically vary drastically from one to the next. One can derive the scaling of the average $\mathcal{G}$ in the thermal regime to be $\mathcal{G}\propto \log|\mathcal{H}|$ and in the localized regime to be $\mathcal{G}\propto -\log|\mathcal{H}|$, leading to the natural definition of the location of the crossover to be the point where $\mathcal{G} = \mathrm{const}$\cite{serbynCriterionManybodyLocalizationdelocalization2015}.

\begin{figure}
    \includegraphics[width=0.45\textwidth]{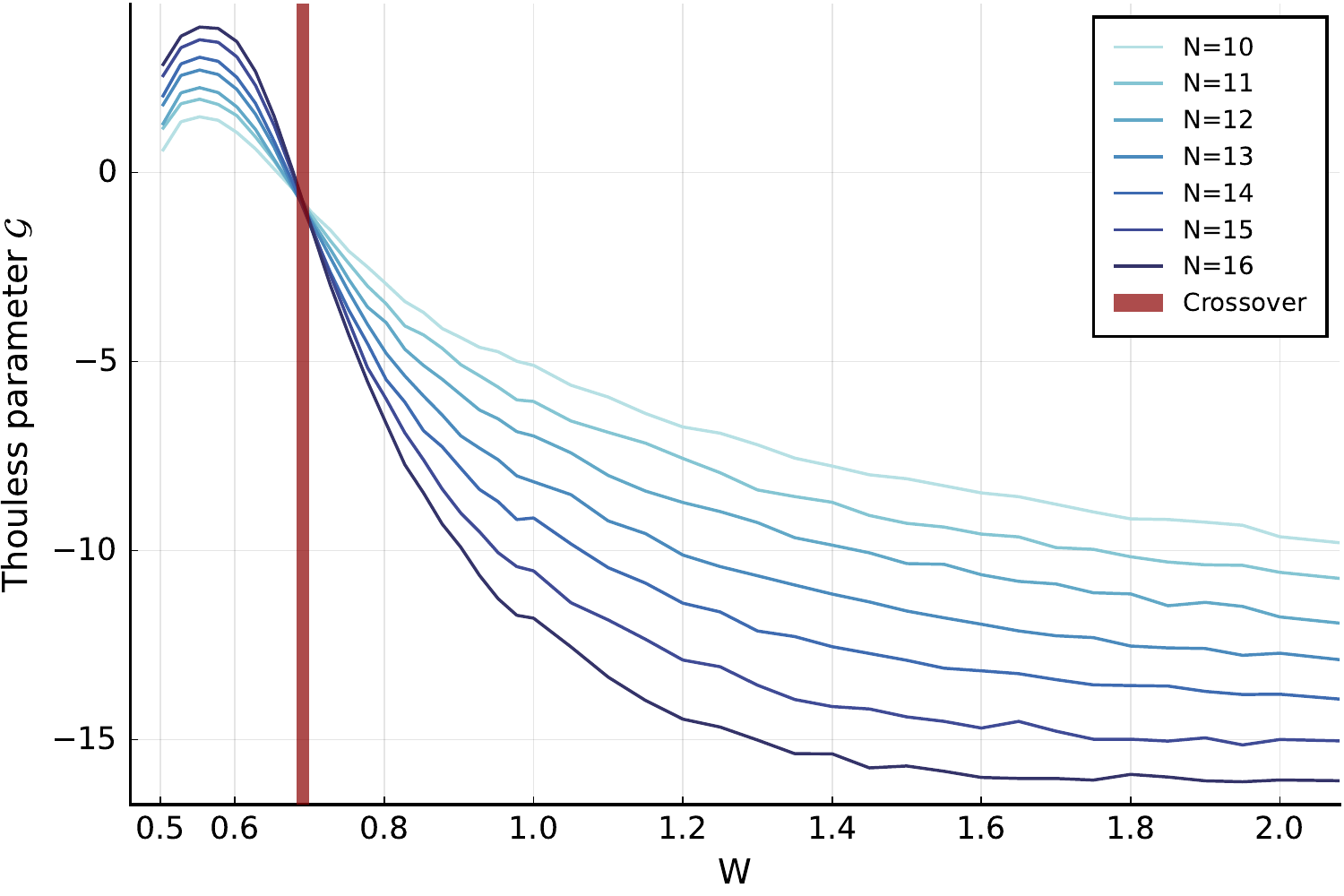}
    \caption{\textbf{Thouless parameter.} Spectral and disorder averaged $\mathcal{G}$ versus disorder strength $W$. Data shown uses local operator $\hat{V}_1=2\hat{S}_z^{(1)}$.}
    \label{fig:thouless_parameter}
\end{figure}

Figure~\ref{fig:thouless_parameter} shows results using local operator $\hat{V}_1=2\hat{S}_z^{(1)}$. Data for local operators $\hat{V}_2 = 4\hat{S}_z^{(1)}\hat{S}_z^{(2)}$ and $\hat{V}_3 = \hat{S}_+^{(1)}\hat{S}_-^{(2)} + \mathrm{H.c.}$ is visually identical. There is a very clear point, where all curves intersect each other indicating the crossover's location. To the right of the crossing point in the localized regime, the curves are roughly evenly spaced, reflecting the expectation of $\mathcal{G}\propto -\log|\mathcal{H}|$, clearly signaling the localized regime. The apparent absence of a drift of the transition point with system size is in contrast to observations in power-law interacting models with on-site disorder and will be further discussed in the next subsection.

\subsection{Half-chain entropy}
Having shown, that there is indeed a localization crossover, we now demonstrate that our effective pair model is indeed a good approximation. We start by probing the half-chain entropy, $S=-\Tr \rho_A \log_2\rho_A$, with $\rho_A=\Tr_B(\rho)$, i.e. the entanglement entropy between two halves of the chain. For that we select $\left\lfloor\frac{N}{2}\right\rfloor$ consecutive spins and trace out the rest, resulting in two cuts due to the periodic boundary conditions, and average over all $N$ possible choices of connected subsystems and all eigenstates.

In an ergodic system, all bulk states should exhibit volume-law entanglement
meaning $S\propto N$. In contrast, in a localized setting all states show area-law entanglement, which for $d=1$ means $S = \mathrm{const}$ \cite{eisertAreaLawsEntanglement2010,gogolinEquilibrationThermalisationEmergence2016}.

To compute the half-chain entropy predicted by the pair model, we need to determine how many pairs are divided by each cut and how often these pairs are found in one of the entangled states $\ket{\pm} = 1/\sqrt{2} (\ket{\uparrow\downarrow} \pm \ket{\downarrow\uparrow})$. Not all pairs consist of adjacent spins [see Fig.~\ref{fig:1}c], so a cut can separate more than one pair. The amount of cut bonds is easily determined from the position data alone, by adding up the distances between paired spins. Respecting periodic boundary conditions of the system yields an additional factor of 2, since there are two cuts needed to divide the chain.

Considering the entropy contribution of a single bond, if we were to average over all possible configurations of pair states, each cut bond would contribute half a bit of entanglement on average, as half of the pair states are maximally entangled and the other half not entangled at all. However, here we consider the sector of smallest positive magnetization, which yields a slightly larger entropy, because it favors the entangled states $\ket{\pm}$ (which have zero net magnetization) over the fully polarized ones. This modification can be computed exactly (see Appendix~\ref{appendix:pair_entropy} for details).

\begin{figure}
    \includegraphics[width=0.45\textwidth]{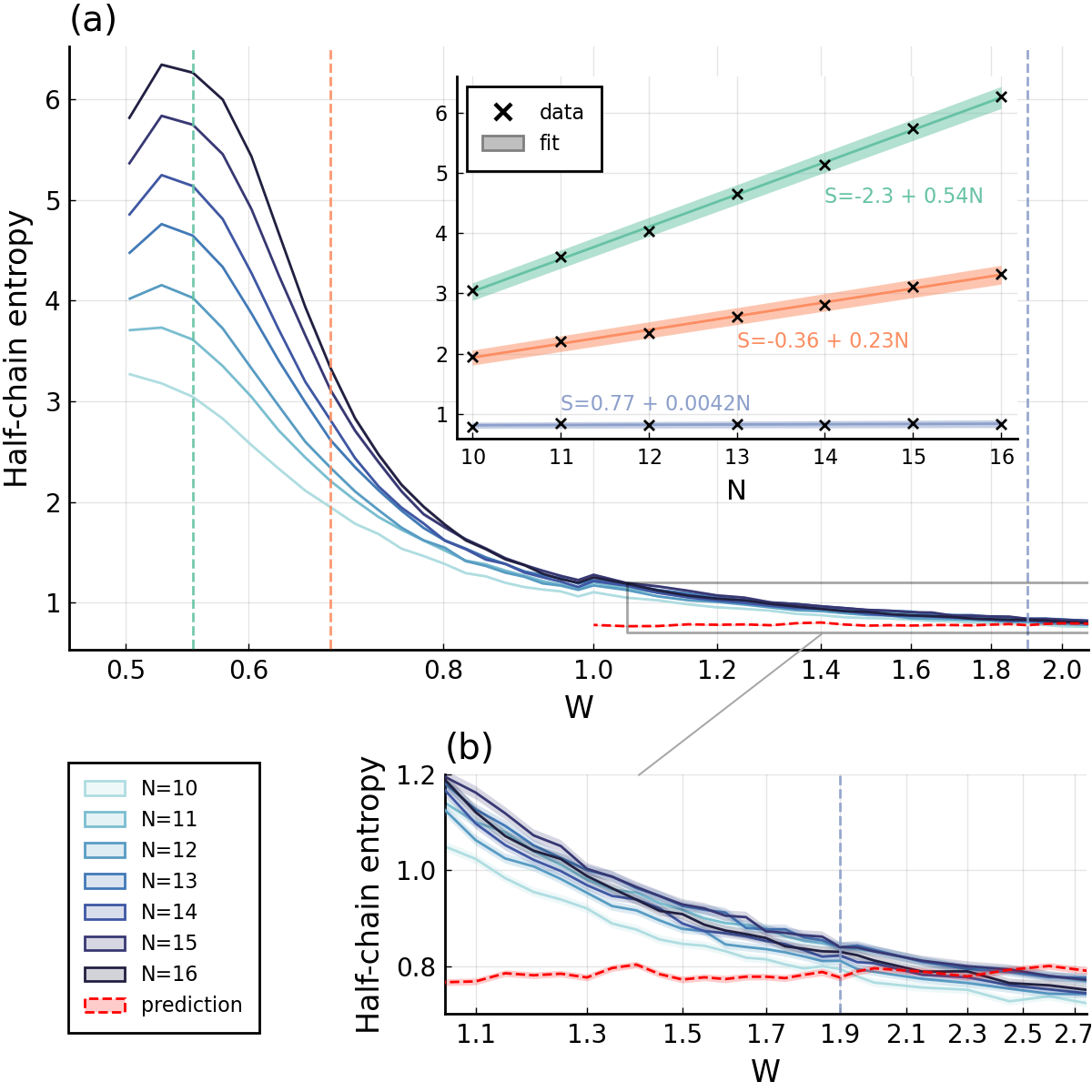}
    \caption{\textbf{Half-chain entropy.} Average over possible cut locations and over disorder realizations for different system sizes as a function of disorder strength. Also shown is the prediction derived from a pair description, computed from position data for $N=16$ (red dashed line), see \ref{appendix:pair_entropy} for details. Inset: Linear fits at fixed disorder strengths indicated by the vertical dashed lines in the main panel. Shaded areas indicate uncertainty from the fit; (b) magnifies the strongly disordered regime of (a). Shaded areas indicate statistical uncertainty from disorder averaging.}
    \label{fig:hce}
\end{figure}

Taking into account both the effects of extended pairs and of the fixed total magnetization, we can compute a prediction for the entanglement entropy directly from the interaction matrix $J_{ij}$. Figure~\ref{fig:hce} shows both the numerically computed values for different system sizes (solid) and pair-model prediction (dashed). 

We clearly see the change between the ergodic and localized regime for the numerically computed data. For strong disorder all lines collapse, confirming on one hand the area law entanglement expected in the localized phase and, on the other hand, validating the pair model as it predicts the strong-disorder limit with high accuracy. Figure~\ref{fig:hce}(b) magnifies the strong-disorder regime showing that the pair-model prediction in fact slightly overestimates the half-chain entropy for very strong disorder. This might indicate, that there are spins that do not pair up perfectly, not forming a maximally entangled Bell pair. It is plausible that this happens at late stages of the pair elimination procedure described in Sec.~\ref{sec:pair_model} when the spins of a pair can have couplings that are stronger than the pair's internal coupling but the spins associated with these stronger couplings are already eliminated. We thus interpret this feature as an indication of the limitations of a simple pair description.

\begin{figure}
    \includegraphics[width=0.45\textwidth]{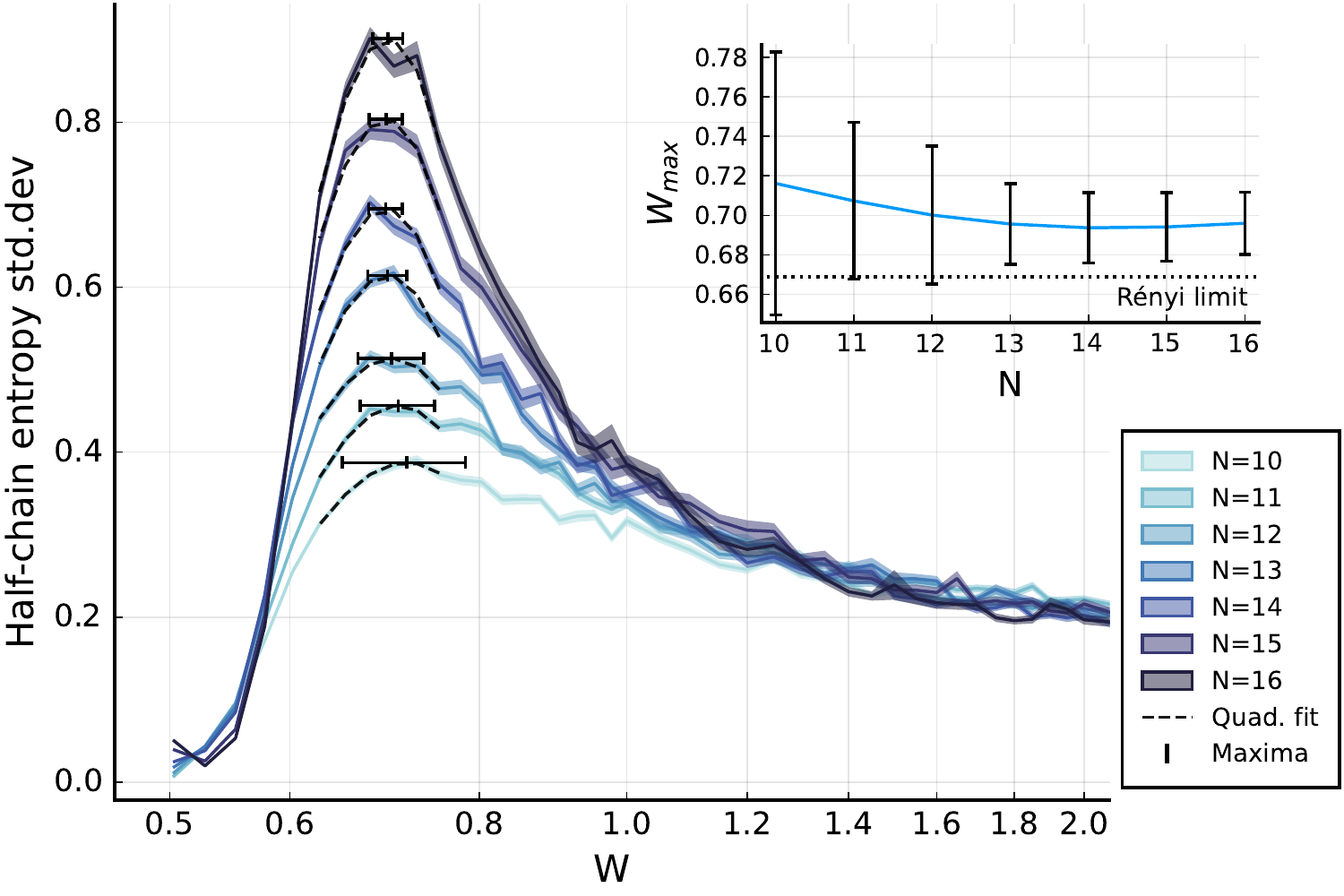}
    \caption{\textbf{Standard deviation of half-chain entropy.} The main plot shows the standard deviation of the half-chain entropy across disorder realizations exhibiting a clear maximum around which a quadratic polynomial is fitted. Shaded areas indicate statistical uncertainty. Inset: Position of the maximum as extracted by the fits. Errors shown are statistical errors from the fits.}
    \label{fig:hce_variance}
\end{figure}

Another piece of information that we can readily access via the half-chain entropy is the location of the crossover. To determine it, we calculate the variance of the half chain entropy over different disorder realizations and extract the maximum for each chain length $N$ via a quadratic fit \cite{kjallManyBodyLocalizationDisordered2014,khemaniCriticalPropertiesManyBody2017}.
Figure~\ref{fig:hce_variance} shows no strong dependence of the crossover point on $N$ in the range of accessible system sizes. Indeed, the crossover does not seem to drift significantly, which is in contrast to models with on-site disorder, see, e.g. Refs.~\cite{khemaniCriticalPropertiesManyBody2017,schifferManybodyLocalizationSpin2019,abaninDistinguishingLocalizationChaos2021}, where  finite-size drifts of the transition point are commonly observed.

Interestingly, the crossover location is very close to the density given by Rényi's parking constant, or jamming limit, which is the maximal density attainable, by randomly placing non-overlapping unit intervals on the number line \cite{renyialfredOneDimensionalProblemConcerning1958}. As in experiments with Rydberg spins, atom positions result from such a random process; this could imply, that these experiments might not be able to reach the densities required for observing the fully ergodic regime. However, it is unclear how the crossover location generalizes to higher dimensions and larger systems. 

\subsection{Participation ratio}
Now that we have seen, that the pair model captures the spatial entanglement structure of the exact eigenstates, we compare the predicted eigenstates directly to the exact ones by computing the participation ratio (PR).
Intuitively, it measures how many states of a reference basis $\mathcal{B}=\{\ket{b}\}$ contribute to a given eigenstate $\ket{\phi_n}$:

\begin{align}
    \mathrm{PR}_\mathcal{B}(\ket{\phi_n}) = \left(\sum_{b\in\mathcal{B}} |\braket{b}{\phi_n}|^4 \right)^{-1} \quad.
\end{align}

Usually, in the MBL context, one chooses a product basis as reference, because a low PR relative to product basis means the eigenstates are close to product states. "Low" in this context means a sublinear scaling of PR with the dimension of the Hilbert space $\mathcal{H}$: $\mathrm{PR} \propto |\mathcal{H}|^{\tau}$, where $\tau < 1$. In contrast, a thermalizing system always has $\mathrm{PR} \propto |\mathcal{H}|$ with respect to any product basis \cite{serbynThoulessEnergyMultifractality2017,maceMultifractalScalingsManyBody2019,luitzMultifractalityItsRole2020}.

Here we compare two different reference bases, the $z$-basis $\mathcal{Z}=\{\ket{\uparrow},\ket{\downarrow}\}^{\otimes N}$ and the pair basis $\mathcal{P}=\{\ket{\pm},\ket{\updownarrow\updownarrow}\}^{\otimes N/2}$, introduced above, to determine how well the pair model describes the eigenstates. If the pair basis $\mathcal{P}$ was exactly equal to the eigenbasis, its PR would be exactly 1. In this case, the expected PR with respect to the z-basis, averaged over the Hilbert space, $\mathcal{Z}$ will be $1.5^{N/2}$, because a single pair has an average PR of $1.5$. However, we only consider the sector of smallest positive magnetization, which increases the expected PR by a similar line of reasoning as for the entropy in the previous section.

\begin{figure}
    \includegraphics[width=0.45\textwidth]{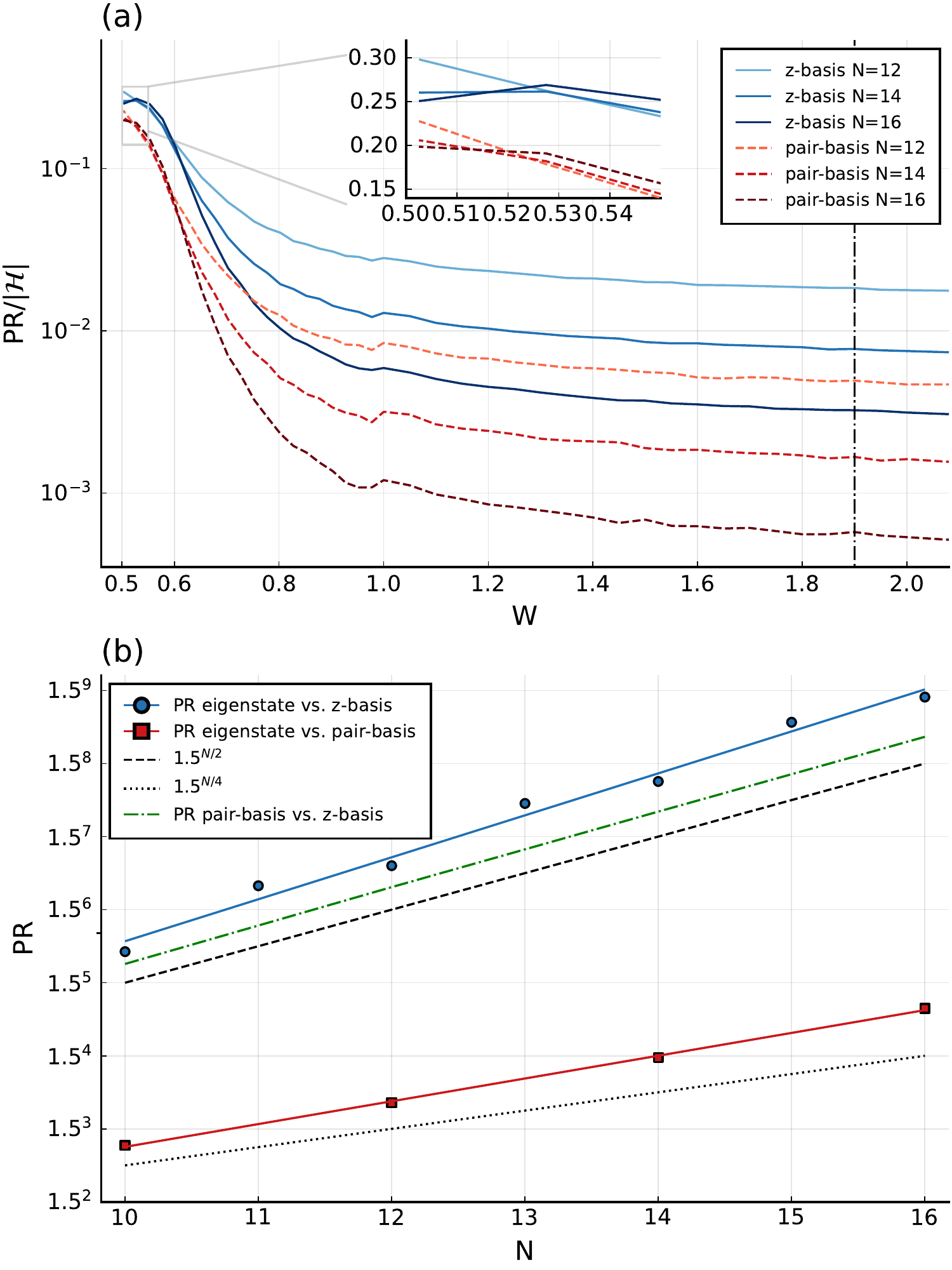}
    \caption{\textbf{Participation ratio.} (a) PR relative to Hilbert space dimension $|\mathcal{H}|$ for different reference bases: $z$-basis in blue, pair basis in red. The inset shows a magnification of the region towards perfectly ordered systems. (b) shows the growth in absolute PR with increasing system size in the localized regime. The used value of $W$ is indicated by the dash-dotted line in (a).}
    \label{fig:pr}
\end{figure}

Figure~\ref{fig:pr}(a) shows the PR relative to the two reference bases as a fraction of the Hilbert space dimension $|\mathcal{H}|$. We see that the weakly disordered regime indeed has ergodic eigenstates as the curves collapse onto each other. 
The small offset between the two reference bases is plausible, since a thermal systems eigenstates express volume law entanglement and thus the overlap with a product basis like $\mathcal{Z}$ is minimal. The states of the pair basis contain pairwise entanglement and are thus a bit closer, which manifest as slightly lower PR. Around $W=0.6$, the scaling with $|\mathcal{H}|$ starts to change to a sublinear relation as we crossover to the localized regime. 

Checking the PR deep in the localized phase (at $W=1.9$) in Fig.~\ref{fig:pr}(b), we can see that the PR relative to the $z$-basis (blue line) is slightly, but systematically, larger than the pair model's prediction (dashed green line). Consistent with this observation, we see that the PR relative to the pair basis (red line), while being much smaller, is still not constant across system sizes. 

We conclude, that the pair states offer a good first order approximation of the true eigenstates, but there are higher order resonances that lead to further hybridization for some states. The exponent of the remaining dependence on system size is close to $N/4$, which hints at effects stemming from interactions between pairs.

\section{Conclusions}
We analyzed a disordered Heisenberg XXZ spin model with power-law interaction and positional disorder, which is naturally realized by many quantum simulation platforms. Among these, cold Rydberg gases allow for easy tuning of the disorder via the sample's density due to the Rydberg blockade. By using standard MBL indicators, we showed numerically that this system undergoes a localization crossover, which we interpreted in terms of a simple physical model derived using an SDRG ansatz. This model, consisting of an effective Ising model of strongly interacting pairs of spins, was verified by considering the PR of eigenstates with the conjectured basis, which is drastically reduced compared to the PR relative to the $z$-basis. Still, there was a weak dependence on system size left, which means there are higher order corrections to our model. Nonetheless, we also showed that this simple model can already  predict the entanglement entropy of the system nearly perfectly.

With this model at hand, we can now make predictions for large systems which may be tested in quantum simulation experiments. Of course, one of the most interesting questions will be, whether the location of the crossover shifts towards stronger disorder for large systems, indicating a transition at infinite disorder strength in the thermodynamic limit. For this purpose the easy tunability of the disorder is a great advantage as both sides of the crossover can be probed on the same platform by changing the system parameters. Remarkably, our small-scale numerical study showed almost no finite-size drift. This could indicate that localization in this model is more stable than in similar models against resonances. We leave this investigation for future work.

Note that the pair model cannot be used to predict the crossover itself as it essentially requires the assumption that one can find strongly interacting pairs, which is only justified in the strongly disordered regime. Recent arguments for the absence of localization postulate the existence of rare thermal subregions within the system \cite{deroeckStabilityInstabilityDelocalization2017,luitzHowSmallQuantum2017,morningstarAvalanchesManybodyResonances2022,selsDynamicalObstructionLocalization2021,selsThermalizationDiluteImpurities2022,selsBathinducedDelocalizationInteracting2022,thieryManyBodyDelocalizationQuantum2018,deroeckStabilityInstabilityDelocalization2017,ponteThermalInclusionsHow2017,pandeyAdiabaticEigenstateDeformations2020}. This would of course break the base assumption of the pair model. A possible direction for future research would be to extend the model to include not only pairs but also larger clusters, which would require one to track all kinds of interactions between clusters of different sizes.

Interestingly, the dimensionality of the system does not directly influence the pair model. As long as the couplings are sufficiently disordered, such that pairs can be defined, it will be a good approximation. Thus it suffices to study how the distribution of couplings changes with respect to the dimensionality $d$ of the space and coupling power $\alpha$. Similar to resonance counting arguments arguments \cite{yaoManyBodyLocalizationDipolar2014}, we conjecture the requirement $d<\alpha$ for the pair model to be applicable. Hence, we expect our results, while acquired in $d=1$, to generalize well to $d>1$.

\section{Acknowledgements}
We thank Dima Abanin for his stimulating input. For numerical simulations we used the Julia programming language \cite{bezansonJuliaFreshApproach2017}. The authors acknowledge support by the state of Baden-Württemberg through bwHPC and the German Research Foundation (DFG) through grant no INST 40/575-1 FUGG (JUSTUS 2 cluster). This paper is supported by the Deutsche Forschungsgemeinschaft (DFG, German Research Foundation) under Germany’s Excellence Strategy EXC No. 2181/1-390900948 (the Heidelberg STRUCTURES Excellence Cluster) and under SFB 1225 ISOQUANT No. 273811115.

\appendix
\section{Derivation of pair picture}\label{appendix:pair_model}
Here we derive the pair model of the main text by means of Schrieffer-Wolff transformations \cite{bravyiSchriefferWolffTransformation2011}. Starting with the full Hamiltonian of the system

\begin{equation}\label{eq:appendix_HeisenbergXXZ}
	\hat{H} = \frac{1}{2}\sum_{i\neq j} J_{ij} \underbrace{ \left( \hat{S}_x^{(i)}\hat{S}_x^{(j)} + \hat{S}_y^{(i)} \hat{S}_y^{(j)} + \Delta \hat{S}_z^{(i)} \hat{S}_z^{(j)} \right) }_{\equiv H\ind{pair}^{(i)(j)}}  \ .
\end{equation}
Suppose without loss of generality that $J_{12}\gg J_{1j},J_{2j}$ and set $H_0 = J_{12}H\ind{pair}^{(1)(2)}$ and $V = H_{XXZ}-H_0$. We label the eigenvectors and eigenenergies of $H\ind{pair}$ like:

\begin{center}
	\begin{tabular}{c|c|c}
		State $k$ & Energy $E_k$ & Vector $\ket{k}$\\\hline
		1 & $2-\Delta$ & $\sqrt{2}^{-1} \left( \ket{\uparrow\downarrow}+\ket{\downarrow\uparrow} \right)$\\
		2 & $\Delta$ & $\ket{\uparrow\uparrow}$\\
		3 & $\Delta$ & $\ket{\downarrow\downarrow}$\\
		4 & $-2-\Delta$ & $\sqrt{2}^{-1} \left( \ket{\uparrow\downarrow}-\ket{\downarrow\uparrow} \right)$
	\end{tabular}
\end{center}

The projectors on these states are consequently named $P_k = \dyad{k}{k}\otimes \mathbb{1}$, but since the middle two states are degenerate, we need to use the projector on the full eigenspace and call it $P_{23}=P_2+P_3$.

To first order only diagonal terms $P_k V P_k$ contribute, which in this case means the pair decouples and only an effective Ising term remains:

\begin{align}
    \hat{H} &= \sum_{i,j}J_{ij} \hat{H}\ind{pair}^{(i)(j)} \\
    &\approx J_{12}\hat{H}\ind{pair}^{(1)(2)} + \sum_{i,j > 2} J_{ij}\hat{H}\ind{pair}^{(i)(j)}\notag\\
    &\quad+ \hat{S}_z^{(1)(2)} \sum_{i>2} \tilde{\Delta}_i \hat{S}_z^{(i)} + \mathcal{O}(\hat{V}^2)
\end{align}
where $2\hat{S}_z^{(1)(2)} = \dyad{\uparrow\uparrow} - \dyad{\downarrow\downarrow}$ is akin to a spin\nobreakdash-1 magnetization operator and $\tilde{\Delta}_i = \Delta (J_{1i}+ J_{2i})$ is the renormalized Ising coupling. Note, that this first order term lifts the apparent degeneracy of the $\ket{\uparrow\uparrow}$ and $\ket{\downarrow\downarrow}$ states. This elimination is a good approximation, if the interaction within the pair is much stronger than any other interaction between a spin of the pair and some other spin.

We can now repeat this elimination step with remaining spins by incorporating the effective Ising terms into $V$. This is justified because its coupling is small and it is already first order perturbation theory and thus including it into the zeroth order of the next pair would mix expansion orders inconsistently.

Further eliminations, now generate effective Ising terms between the states $\ket{\uparrow\uparrow}$ and $\ket{\downarrow\downarrow}$ of the eliminated pairs. After pairing up all spins, we find
\begin{align}
    \hat{H} &= \sum_{i,j}J_{ij} \hat{H}\ind{pair}^{(i)(j)}\\
    &\approx \sum_{\langle i,j\rangle} J_{ij}\hat{H}^{(i)(j)}\ind{pair} + \sum_{\langle i,j\rangle, \langle i',j'\rangle} \tilde{\Delta}_{(i,j),(i',j')} \hat{S}_z^{(i)(j)}\hat{S}_z^{(i')(j')}
\end{align}
where the sum over $\langle i,j\rangle$ denotes pairs of spins and $\tilde{\Delta}_{(i,j),(i',j')} = \Delta (J_{i,i'} + J_{j,i'} + J_{i,j'} + J_{j,j'})$.

Also note that with each elimination step, the mean inter-particle distance grows and thus the disorder in the system increases \cite{fisherRandomAntiferromagneticQuantum1994, vasseurQuantumCriticalityHot2015} making it more likely for later elimination steps to be good approximations.

\section{Pair entropy in a specific magnetization sector}\label{appendix:pair_entropy}
Averaged over all states, each cut separating a pair gives an average entropy of $\frac{1}{2}$, since two of the pair's eigenstates are fully entangled and the other two possess no entanglement. However, when we consider a sector of fixed magnetization, this simple argument no longer holds as there are now dependencies among the eigenstates given by the external constraint. Sectors around zero magnetization will have more entropy on average and strongly magnetized sectors less, simply because the strongest magnetized eigenstates possess no entropy.

Given $N$ the number pairs of spins where $N_+$, $N_-$, and $N_0$ pairs occupy the states $\ket{\uparrow\uparrow}$, $\ket{\downarrow\downarrow}$, and $\ket{\uparrow\downarrow}\pm\ket{\downarrow\uparrow}$, we find the number of possible configuration with these amounts to be
\begin{equation}
    C(N_+,N_-,N_0) = \binom{N}{N_0}\binom{N-N_0}{N_+}2^{N_0}
\end{equation}

In the end we need the number of configurations $\mathcal{C}(N,r) = \sum_{N_0}\mathcal{C}(N,r,N_0) $ given a total amount of pairs $N$ and an magnetization imbalance $r=N_+ - N_-$, where 
\begin{equation}
    \mathcal{C}(N,r,N_0) = \sum_{0 \leq N_+, N_-} C(N_+,N_-,N_0) \delta_{N,N_+ + N_- + N_0} \delta_{r,N_+ - N_-}.
\end{equation}

To evaluate this expression, we compute the generating function
\begin{align}
    \mathcal{Z}(x,y,z) &= \sum_{N>0} x^N \sum_{-N \leq r \leq N} y^r \sum_{N_0 > 0} z^{N_0} \mathcal{C}(N,r,N_0)\\
    &=\smashoperator{\sum_{0\leq N_+,N_0,N_-}}\!\! x^{N_++N_0+N_-} y^{N_+-N_-}z^{N_0}C(N_+,N_-,N_0)\\
    &= \sum_{\mathclap{0\leq N_-}} \! \left(\frac{x}{y}\right)^{N_-}\!\!\! \sum_{\mathclap{0\leq N_+}} (xy)^{N_+} \binom{\!N_+\!+\!N_-\!}{N_+} \sum_{N_0} \binom{\!N\!}{\!N_0\!} (2z)^{N_0}\\
    &= \frac{y}{y-2xyz-xy^2-x}
\end{align}
where we used the fact that $(1-x)^{-k-1} = \sum_n \binom{n+k}{k}x^n$ twice and then a geometric series.

From that, it follows directly that
\begin{align}
    \mathcal{Z}(x,y,1) &= \sum_{N>0} x^N \sum_{-N \leq r \leq N} y^r \mathcal{C}(N,r)\\
    &= \frac{y}{y-2xy-xy^2-x}\\
    &= \frac{1}{1-x\frac{(y+1)^2}{y}}\\
    &= \sum_{0 \leq k} x^k \left(\frac{(y+1)^2}{y}\right)^k\\
    &= \sum_{0 \leq k} x^k \sum_{0 \leq l \leq 2k}y^{l-k}\binom{2k}{l}
\end{align}
and thus by identification of terms
\begin{equation}
    \mathcal{C}(N,r) = \binom{2N}{r+N}\qquad.
\end{equation}

Singling out a specific pair and asking how often it is in one of the entangled states given a set of configurations described by values for $(N_+, N_0, N_-)$, we find that it's the case in 
\begin{equation}
    S(N_+,N_-,N_0) = 2 C(N_+,N_-,N_0-1) = \frac{N_0}{N}C(N_+,N_-,N_0)
\end{equation}
configurations. Again we want to find this number for a total amount of pairs $N$ and an magnetization imbalance $r=N_+ - N_-$. Fortunately, we can find the generating function $\mathcal{Z}_\mathcal{S}(x,y,z)$ of $\mathcal{S}(N,r,N_0) = \frac{N_0}{N}C(N,r,N_0)$ by means of $\mathcal{Z}$:
\begin{equation}
    \mathcal{Z}_\mathcal{S}(x,y,z) = \int \frac{\dd x}{x} z\pdv{z} \mathcal{Z}(x,y,z)
\end{equation}

So we compute:
\begin{align}
    \mathcal{Z}_\mathcal{S}(x,y,z=1) &= \sum_N x^N \sum_r y^r \mathcal{S}(N,r)\\
    &= \int \frac{\dd x}{x} \frac{2xy^2}{(y-x(y+1))^2}\\
    &= \frac{2y^2}{(y+1)^2}\frac{1}{y-x(y+1)^2}\\
    &= 2 \sum_k x^k \sum_l y^{l-k+1}\binom{2k-2}{l}\\
    \Rightarrow \mathcal{S}(N,r) &= 2 \binom{2N-2}{r+N-1}
\end{align}

Thus, cutting a single pair contributes 
\begin{align}
    \bar{S}(N,r) &= \frac{\mathcal{S}(N,r)}{\mathcal{C}(N,r)}\\
    &= 2\frac{N^2-r^2}{4N^2-2N}
\end{align}
bits of entropy, on average over all states in a given magnetization sector.

\begin{figure}
    \includegraphics[width=0.45\textwidth]{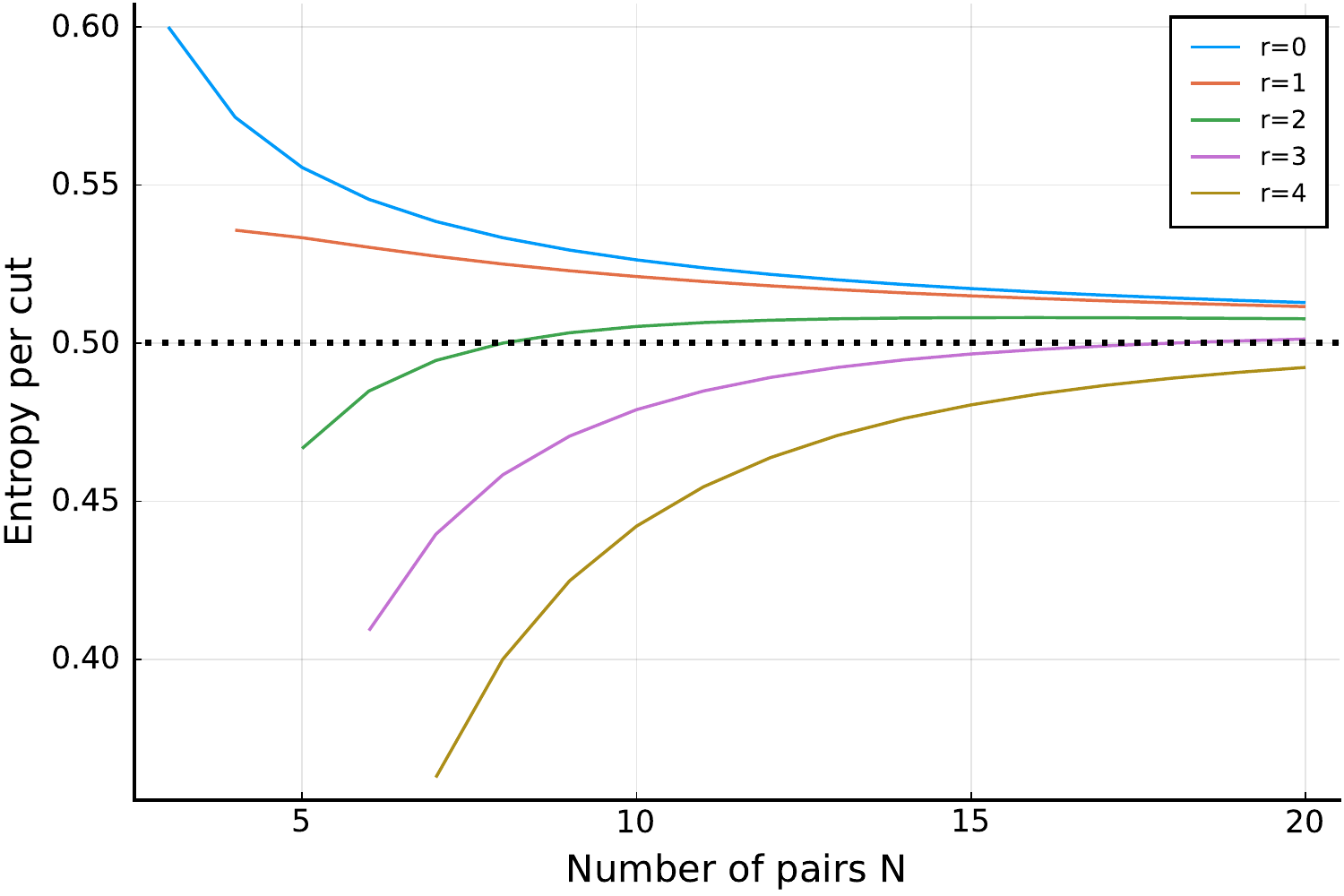}
    \caption{Entropy value of a single cut for different magnetization sectors.}
    \label{appendixfig:entropy_per_cut}
\end{figure}

For the prediction of the average entropy in Fig.~\ref{fig:hce}, we extracted the size of the pairs from the position data, which directly determines how many times a pair is cut, when moving along the chain. The number of cut pairs is then divided by the number of cuts made -- which is equal to the number of spins -- and multiplied by the average entropy contributed by cutting a pair, computed here.

\section{Drawing blockaded positions}\label{appendix:drawing_positions}

In the following, we restrict ourselves to $N$ spins in $d=1$ dimension and measure every distance in units of the blockade radius $r_b$. We define the density of spins $0 \leq \rho=\frac{1}{2W} \leq 1$, the corresponding volume of the space $L = \frac{N}{\rho}$ and set out to construct a scheme to efficiently generate a set of independently drawn positions $\{x_i\}$, that respect the blockade condition
\begin{equation}\label{app:eq:blockade_condition}
    |x_i - x_j| \geq r_b\qquad \forall i\neq j.
\end{equation}

A priori, all positions are drawn i.i.d. from a uniform distribution over the full space $\mathcal{U}[0,L]$ and the naive way would be to just draw $N$ positions and reject the sample if the blockade condition (Eq.~\ref{app:eq:blockade_condition}) is violated. This is essentially equivalent to a random sequential adsorption process where the expected density in $d=1$ is given by Renyi's parking constant $m\approx0.748$ \cite{renyialfredOneDimensionalProblemConcerning1958}. It directly follows, that the rejection rate will become essentially $1$ for any $\rho > m$ and we certainly will not get close to the fully ordered regime. 

To circumvent this problem, we parameterize the positions like
\begin{equation}
    x_i = is + \sigma_i,
\end{equation}
where $s=\frac{1}{\rho}=2W$ is the mean inter-spin distance and $\sigma_i \sim \mathcal{U}[-\sigma,\sigma]$ are i.i.d. random variables. For $\sigma=\frac{L}{2}$ this ansatz is certainly equivalent to the naive scheme. 

Note that, in the highly ordered case $\rho=1-\epsilon$, where $\epsilon$ is small, each realization of the experiment looks essentially like a regularly spaced chain with $s=\frac{1}{1-\epsilon}\approx r_b(1+\epsilon)$ where each site has small fluctuations around the mean. This means, in this limit we get away with choosing $\sigma \approx \epsilon$.

For our simulations, we used the just described method in the region $W < 1.0$ and chose $\sigma = 1.5(\frac{1}{\rho} - 1)$. For $W\geq 1.0$, we used the naive sampling strategy. One can see a slight jump in all plots at $W=1.0$ where the sampling method changes.

\bibliography{references}

\end{document}